# Observation of interlayer phonon modes in van der Waals heterostructures

Chun Hung Lui[1], Zhipeng Ye[2], Chao Ji[2], Kuan-Chang Chiu[3], Cheng-Tse Chou[3], Trond I. Andersen[1], Casie Means-Shively[2], Heidi Anderson[2], Jenn-Ming Wu[3], Tim Kidd[2], Yi-Hsien Lee[3]*, and Rui He[2]*

[1] *Department of Physics, Massachusetts Institute of Technology, Cambridge, Massachusetts 02139, USA*
[2] *Department of Physics, University of Northern Iowa, Cedar Falls, Iowa 50614, USA*
[3] *Department of Materials Science and Engineering, National Tsing Hua University, Hsinchu 30013, Taiwan*
*Corresponding authors (emails: yhlee.mse@mx.nthu.edu.tw, rui.he@uni.edu)

**Abstract:** We have investigated the vibrational properties of van der Waals heterostructures of monolayer transition metal dichalcogenides (TMDs), specifically $MoS_2$/$WSe_2$ and $MoSe_2$/$MoS_2$ heterobilayers as well as twisted $MoS_2$ bilayers, by means of ultralow-frequency Raman spectroscopy. We discovered Raman features (at 30 ~ 40 $cm^{-1}$) that arise from the layer-breathing mode (LBM) vibrations between the two incommensurate TMD monolayers in these structures. The LBM Raman intensity correlates strongly with the suppression of photoluminescence that arises from interlayer charge transfer. The LBM is generated only in bilayer areas with direct layer-layer contact and atomically clean interface. Its frequency also evolves systematically with the relative orientation between of the two layers. Our research demonstrates that LBM can serve as a sensitive probe to the interface environment and interlayer interactions in van der Waals materials.

Two dimensional (2D) atomic crystals, such as graphene, hexagonal boron nitride (hBN) and transition metal dichalcogenides (TMDs), e.g. $MoS_2$, $MoSe_2$, $WS_2$, and $WSe_2$, have risen as a new generation of materials with remarkable properties [1,2]. With the rapid development of sample growth, characterization and fundamental studies of individual 2D materials, recent research frontier advances to explore their hybrid systems. In particular, their flat and inert surfaces enable us to produce stacks of different 2D crystals with atomically sharp interfaces, coupled vertically only by van der Waals forces [3]. In these heterostructures, interactions between two atomic layers can dramatically change the properties of the system and induce phenomena that are absent in individual layers. Recent experiments have, for instance, demonstrated emergent massive Dirac fermions and Hofstadter butterfly in graphene/hBN superlattices [4-6], and the formation of interlayer excitons in TMD heterobilayers [7-11]. The (opto)electronic devices of 2D layered heterostructures also exhibit performance superior to that of traditional devices with lateral 2D junctions [12-18].

While much research has been directed to explore the novel electronic properties of 2D van der Waals heterostructures, it is also important to study their vibrational properties that may affect device performance through electron-phonon interactions. In particular, it is intriguing to explore the possibility of generating new phonon modes through hybridization of different 2D crystals. Recent research [19-31] has revealed that interlayer interactions in few-layer graphene and TMDs can create a set of shear modes and layer-breathing modes (LBMs) that involve lateral and vertical displacement of individual layers, respectively. These interlayer phonon modes provide sensitive probes to layer thickness [19-31], stacking order [23-25, 32] and surface adsorbates [26] of 2D materials. In 2D heterostructures, however, no such phonons have been reported thus far. Indeed, realization of well-defined interlayer phonons is expected to be challenging in 2D heterostructures due to the generally uncontrollable interfacial environment in these systems.

In this Letter, we report the first observation of interlayer phonon modes in atomically thin van der Waals heterostructures. We measured the low-frequency Raman response of $MoS_2/WSe_2$ and $MoSe_2/MoS_2$ heterobilayers (Figure 1a-c). We discovered a distinctive Raman mode (30 - 35 $cm^{-1}$) that cannot be found in any individual monolayers. By comparing with Raman spectra of bilayer (2L) $MoS_2$, 2L $MoSe_2$ and 2L $WSe_2$, we identified the new Raman mode as the LBM arising from the perpendicular vibration between the two rigid TMD layers. We found that the heterogeneous LBM Raman intensity correlates strongly with the suppression of photoluminescence (PL) that arises from interlayer charge transfer. The LBM only emerges in bilayer regions with atomically close layer-layer proximity and clean interface. In addition, the LBM frequency exhibits noticeable dependence on the relative orientation between the two TMD layers, which implies a change of interlayer separation and interlayer coupling strength with the layer stacking. Our results reveal that LBM generally exists in van der Waals heterostructures and can serve as an effective probe to the interface environment and interlayer interactions in these materials.

Our TMD heterostructures were constructed from $MoS_2$, $MoSe_2$ and $WSe_2$ monolayers grown by chemical vapor deposition (CVD) on oxidized silicon substrates [33, 34]. The single-crystal flakes with single-layer (1L) thickness were identified by their optical contrast and characteristic triangular shape, and further confirmed by Raman, PL and atomic-force-microscopy (AFM) measurements. The methods of sample growth and characterization can be found in the literature [33, 34]. We deposited PMMA on the TMD flakes and peeled them off from the substrate in KOH solution at $T < 60$ ℃. Afterward, we transferred the PMMA/TMD layer onto another substrate with as-grown TMD samples, and then removed the PMMA by acetone solution. Due to the numerous TMD flakes on each substrate, we were able to obtain dozens of heterobilayer samples with various stacking order in one transfer. To further clean the sample surface and interface, we annealed the samples at $T = 150$ ºC and low pressure (P = 0.05 Torr) in the flow of argon (20 sccm) and hydrogen (2 sccm) gas for two hours. From the PL and Raman measurements, we confirmed the good quality of TMD layers and interfaces in most of our samples.

We measured the PL and Raman response of our samples using a commercial Horiba Labram HR micro-Raman system in ambient conditions. The Raman setup offers access to frequencies down to 10 $cm^{-1}$ and spectral resolution of ~0.5 $cm^{-1}$. The excitation laser (wavelength 532 nm) was focused onto the samples with spot diameter of ~1 μm and incident power of ~1 mW. We obtained PL and Raman spatial maps by raster scanning with 0.3 μm step size using a precision 2D mapping stage.

Figure 1d-e display the representative PL spectra (in a logarithmic scale) of $MoS_2/WSe_2$ and $MoSe_2/MoS_2$ bilayer stacks, in comparison with the spectra of monolayer areas of the same samples. All 1L $MoS_2$, 1L $WSe_2$, and 1L $MoSe_2$ spectra exhibit strong PL signals, with their characteristic emission energies at ~1.85 eV, ~1.59 eV and ~1.56 eV, respectively [35-37]. The observed PL comes from the recombination of excitons across the direct band gap at the K and K' valleys of their respective band structure [35-37]. Remarkably, the PL intensity is strongly reduced in the overlapping $MoS_2/WSe_2$ and $MoSe_2/MoS_2$ regions. The reduction of PL intensity can vary from a few tens of percent to two orders of magnitude for different heterobilayer samples due to their different quality. Similar PL quenching has also been observed in various TMD heterostructures in prior studies [7-11]. The underlying mechanism is known to be the interlayer charge transfer. Previous research has calculated the electronic structure of different TMD heterostructures [38-46]. The conduction and valence band edges in 1L $MoS_2$ are predicted to be lower than those in 1L $WSe_2$ and 1L $MoSe_2$, giving rise to type II band alignment with staggered gap in their heterojunction [38, 39] (Figure 1f). The photoexcited electrons in the $WSe_2$ ($MoSe_2$) layer tend to flow to the $MoS_2$ layer, and the holes in the $MoS_2$ layer tend to flow to the $WSe_2$ ($MoSe_2$) layer. The spatial separation of the electrons and holes therefore suppresses the intralayer optical recombination processes and thus the PL intensity in each individual layer. The strong PL suppression (up to two orders of magnitude) observed in our samples indicates that the charge transfer rate is close to the rate of exciton generation under our continuous laser excitation. Such high transfer efficiency is consistent with the remarkable transfer time scale (<50 fs) determined by ultrafast studies [11].

We have also measured the Raman response of the heterobilayer samples. As shown in Figure 2, a pronounced Raman peak emerges at ~32 $cm^{-1}$ for both $MoS_2/WSe_2$ and $MoSe_2/MoS_2$ heterobilayers. This Raman feature is not observed in any TMD monolayers (Figure 2a,i), indicating that it arises from the interaction between the two TMD layers. The same Raman feature has been observed in more than 70 heterobilayer samples, with frequencies varying between 30 and 35 $cm^{-1}$ for both types of heterobilayers (we will discuss the sample dependence in more details later in this paper).

To understand the origin of the new Raman mode, we have measured the low-frequency Raman spectra of Bernal-stacked 2L $MoS_2$, 2L $WSe_2$ and 2L $MoSe_2$ samples (Figure 2c-h). All these bilayer spectra exhibit two pronounced Raman peaks. Based on the reported low-frequency modes in Bernal 2L $MoS_2$ and 2L $WSe_2$ [27-31], we assign the low-frequency peaks (17 - 22 $cm^{-1}$) in all the three samples as the interlayer shear mode and the high-frequency peaks (30 - 40 $cm^{-1}$) as the LBM. The line width of these interlayer modes generally increases with the increase of mode frequency. For instance, the full width at half maximum (fwhm) of the LBM increases from 3.5 to 9.5 $cm^{-1}$ as its frequency increases from 2L $WSe_2$ (29.5 $cm^{-1}$) to 2L $MoSe_2$ (30.5 $cm^{-1}$) and to 2L $MoS_2$ (39.5 $cm^{-1}$), and the LBM is also broader than the shear mode (Table 1). These observations reflect the increasing anharmonic decay rate of phonons at higher energy. From comparison with the heterobilayer spectra, we find that the frequency (~32 $cm^{-1}$) of the new Raman mode in the $MoS_2/WSe_2$ ($MoSe_2/MoS_2$) heterobilayer lies between the LBM frequencies of 2L MoS$_2$ and 2L WSe$_2$ (2L MoSe$_2$) (vertical dashed lines in Figure 2). Also, the width of the new Raman mode (fwhm = 8 cm$^{-1}$ and 5.5 cm$^{-1}$ for MoS$_2$/WSe$_2$ and MoSe$_2$/MoS$_2$ heterobilayers, respectively) roughly matches those of the LBMs from Bernal bilayers (Table 1). Therefore, our observations strongly suggest that the new Raman mode in the heterobilayers originates from the LBM vibration between the two TMD layers (Figure 2b,j).

The new Raman mode has been observed in heterobilayers with different stacking order. This indicates that the formation of LBM is robust against the change of relative orientation (translation and rotation) between the two TMD layers. In contrast, the TMD heterobilayers do not exhibit any Raman feature near the shear-mode range (17 - 22 cm$^{-1}$) for Bernal bilayers, indicating the absence of shear mode in the heterostructures.

We can understand the above observations by considering the interlayer interactions in the heterostructures. As the MoS$_2$ and WSe$_2$ (MoSe$_2$) monolayers have ~4% mismatch of lattice constant and orient randomly with one another in the heterobilayers, the two lattices are generally incommensurate. Thus lateral displacement of the two layers does not produce any overall restoring force, and hence no shear mode vibration can be generated in the heterostructures. Vertical displacement of the two layers, however, can create a finite restoring force due to the van der Waals interactions between them. This gives rise to the LBM vibration. As the overall strength of the van der Waals force depends predominantly on the interlayer distance, the formation of LBM vibration should be stable regardless of the detailed lattice matching in the lateral dimension. This is supported by the similar LBM observed in trilayer graphene with ABA and ABC stacking order [25].

For a more quantitative understanding, we treat the TMD bilayer as two spring-coupled masses with frequency $\omega = \sqrt{k/m}$, where $m$ and $k$ are the reduced mass (per unit area) and effective force constant of the bilayer system, respectively. For Bernal-stacked 2L MoS$_2$ and 2L WSe$_2$, we obtain $m_{2L-WSe_2} = 1.97\, m_{2L-MoS_2}$ from their mass densities, and $\omega_{2L-WSe_2} = 0.75\, \omega_{2L-MoS_2}$ from their measured LBM frequencies. From these relations, we find that $k_{2L-WSe_2} = 1.11\, k_{2L-MoS_2}$, indicating that their force constants are nearly equal. By simply using an average force constant and mass density in the MoS$_2$/WSe$_2$ heterobilayer, we obtain $\omega_{MoS_2/WSe_2} = 0.89\, \omega_{2L-MoS_2} = 35$ cm$^{-1}$. Similarly, we estimate the LBM frequency of the MoSe$_2$/MoS$_2$ heterobilayer to be 34.5 cm$^{-1}$. Both predicted LBM frequencies are very close to the frequencies of the observed new Raman mode (30 - 35 cm$^{-1}$), and further support our LBM assignment. The overall lower LBM frequency observed in the experiment suggests weaker interlayer coupling in the heterostructures than in the Bernal bilayers.

To further support our arguments above, we have also examined twisted MoS$_2$ bilayers, which are formed by randomly stacking two MoS$_2$ monolayers together. Figure 2e displays a representative Raman spectrum, which exhibits a pronounced Raman peak at 35 cm$^{-1}$ with fwhm = 6.5 cm$^{-1}$. The same Raman mode has been observed in more than 60 twisted 2L MoS$_2$ samples, with frequencies varying between 32 and 39 cm$^{-1}$. These results are compatible to the LBM in Bernal 2L MoS$_2$, which has slightly higher frequency (39.5 cm$^{-1}$) and correspondingly larger line width (fwhm = 9.5 cm$^{-1}$) (Table 1). Our results indicate that LBM can be formed in twisted bilayer van der Waals materials [32, 47, 48]. The decreased LBM frequency in twisted bilayers is reasonable because the two layers are packed less efficiently as shown by other studies [47, 48], resulting in weaker interlayer coupling. In addition, the shear mode is quenched in twisted bilayers due to the lateral lattice mismatch. These observations are consistent with our results and interpretation for the MoS$_2$/WSe$_2$ and MoSe$_2$/MoS$_2$ heterostructures.

As the LBM arises directly from the interlayer coupling, it is interesting to investigate its sensitivity to the interface conditions. To this end, we have mapped the spatial profile of the LBM in an unannealed MoS$_2$/WSe$_2$ heterobilayer sample, which contains regions of 1L MoS$_2$, 1L WSe$_2$ and overlapping MoS$_2$/WSe$_2$ layers (Figure 3a). As expected, the LBM is observed only in the area where the two TMD monolayers overlap (Figure 3b). However, the LBM emerges only on the left side of the MoS$_2$/WSe$_2$ region. To understand the inhomogeneous LBM response, we have also mapped the intensity of the PL peaks of 1L MoS$_2$ and 1L WSe$_2$ at ~1.85 and ~1.59 eV, respectively (Figure 3c-d). Both PL intensities decrease in the MoS$_2$/WSe$_2$ region due to the interlayer electron-hole separation. The PL reduction is, however, much more severe on the left side than on the right side, indicating more efficient interlayer charge transfer and hence better layer-layer contact on the left side. This is further confirmed by AFM analysis. The AFM topographic image of the sample exhibits two distinct regions with different heights in the overlapping MoS$_2$/WSe$_2$ area (Figure 3e-f). The left side of the area has a step height of ~0.7 nm (blue lines in Figure 3e-f), which is the same as the monolayer step height measured on exfoliated multilayer MoS$_2$ crystals [49] and CVD MoS$_2$ grown on SiO$_2$ substrates [34]. The two TMD layers are therefore in direct contact in this region. In contrast, the right side has a larger interlayer separation of ~1.8 nm (green lines in Figure 3e-f), which can be attributed to unintentional residues trapped at the interface. The overall agreement between the LBM/PL intensity mapping and AFM scanning shows that the interlayer phonons can only be generated in heterostructure regions with direct interlayer contact and atomically clean interface. Our results also indicate that the observed Raman signals are not from any interfacial molecules.

The interface of the heterobilayers can be improved by annealing the samples. Figure 4 shows the optical image, LBM and PL mappings of an annealed MoSe$_2$/MoS$_2$ heterobilayer sample. The LBM and the associated PL reduction are observed in most of the overlapping MoSe$_2$/MoS$_2$ areas, indicating an overall good layer-layer contact in the bilayer region. We also examined the AFM images of multiple annealed samples (not shown) and consistently found good interface conditions.

Finally, we investigated the dependence of LBM on the relative orientation of the layers in the heterostructures. Our TMD monolayer flakes typically exhibit triangular shapes, which have been shown to be single crystals that terminate with the transition-metal atoms (Mo, W) at zigzag edges [50, 51]. Therefore we can conveniently determine the orientation of individual TMD flakes and the rotational angle ($\theta$) between the top and bottom TMD flakes from their optical images (with uncertainty < 5º). The orientations of some TMD flakes were further confirmed by second harmonic generation measurements [47]. In our experiment, we have measured a large collection of $MoSe_2/MoS_2$ heterobilayer samples, among which 58 of them exhibit strong LBM spectra and well-defined rotational angle between the $MoSe_2$ and $MoS_2$ layers. Figure 5a-f display the optical images and LBM spectra of three representative heterobilayers with $\theta \approx 45º$, 32º and 5º. Figure 5g shows the LBM frequency of all 58 samples as a function of $\theta$. While the LBM frequency varies somewhat irregularly due to the different sample conditions (e.g. strain, defects, surface and interfaces), the average value is found to decrease from 33.5 to 30.7 $cm^{-1}$ as $\theta$ increases from 0º to 60º. The result clearly demonstrates the sensitivity of the LBM to the layer stacking. The softening of LBM implies that the interlayer coupling becomes slightly weaker as $\theta$ increases. According to the previous first-principles calculation by J. Kang *et al* [44], $MoSe_2/MoS_2$ heterobilayers are expected to exhibit moiré patterns with a quasi-period (< 10 nm) that varies with $\theta$ due to the 4% lattice mismatch between the $MoSe_2$ and $MoS_2$ layers. In addition, J. Kang *et al* predicted an average interlayer distance $d = 3.345$ Å and interlayer adsorption energy $E_{ad} = 160$ meV for $\theta = 0º$, but a slightly larger interlayer distance $d = 3.368$ Å and smaller adsorption energy $E_{ad} = 157$ meV for $\theta = 60º$ in the $MoSe_2/MoS_2$ heterobilayer [see Figure 1 and Table 1 in Ref. (44), where Pattern A and B correspond to $\theta = 0º$ and 60º, respectively]. These theoretical results agree qualitatively with the observed angle-dependence of LBM in our experiment.

In conclusion, we have observed the layer-breathing mode (LBM) phonons in van der Waals heterostructures formed from transition metal dichalcogenide (TMD) monolayers. The LBM Raman intensity correlates strongly with the PL suppression. The LBM emerges only in bilayer areas with atomically clean interface and exhibits noticeable dependence on the relative orientation of the two layers. Our study demonstrates a nondestructive and convenient method of characterizing functionalized 2D materials, which can be further applied to study interlayer modes in higher frequency range [52] and in heterostructures formed from other 2D crystals [3]. Our research also opens the possibility of fabricating novel nanoscale phononic crystals in a bottom-up approach. By stacking or epitaxially growing different 2D crystals on one another, one may potentially fabricate artificial materials with unprecedented phonon structures and desirable vibrational properties that cannot be realized in natural crystals.

We acknowledge the Donors of the American Chemical Society Petroleum Research Fund (Grant 53401-UNI10) for support of this research. This research is also supported by NSF grant (DMR-1206530, DMR-1410496). R. H. acknowledges support from UNI Faculty Summer Fellowship. Z. Y. acknowledges support from the UNI Physics Department International Student Fund. Y. H. L. acknowledges support from the Ministry of Science and Technology of the Republic of China (103-2112-M-007-001-MY3).

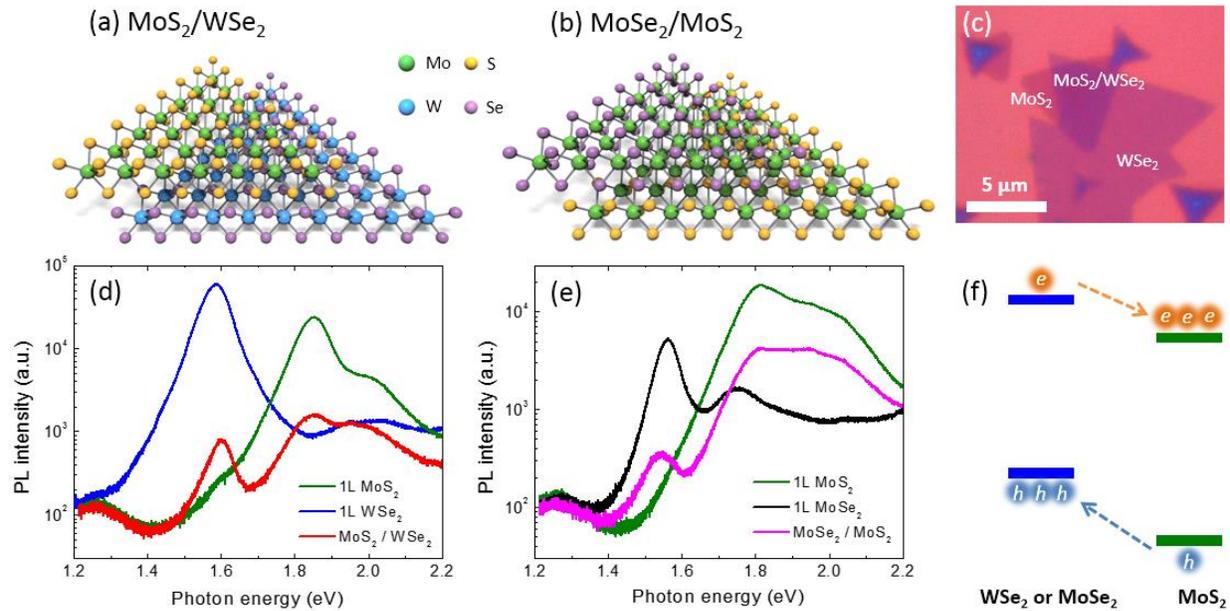

Figure 1. (a-b) Schematic atomic configurations of the MoS$_2$/WSe$_2$ and MoSe$_2$/MoS$_2$ heterobilayers. (c) Optical image of a MS$_2$/WSe$_2$ heterobilayer. (d) Photoluminescence (PL) spectrum of a MoS$_2$/WSe$_2$ heterobilayer in comparison with spectra of the 1L MoS$_2$ and 1L WSe$_2$ regions in the same sample under the same measurement conditions. (e) Similar PL spectra for a MoSe$_2$/MoS$_2$ heterobilayer. The excitation laser wavelength is 532 nm for all spectra. (f) Schematic of interlayer charge transfer in the heterostructure. The blue bars represent the conduction and valence band edges of 1L WSe$_2$ or MoSe$_2$. The green bars represent those of 1L MoS$_2$. They form type II heterostructures with staggered band gap. The photoexcited electrons tend to flow from the WSe$_2$ (MoSe$_2$) layer to the MoS$_2$ layer, and the holes from the MoS$_2$ to WSe$_2$ (MoSe$_2$) layer. The resultant charge separation strongly suppresses the intralayer recombination processes and the PL.

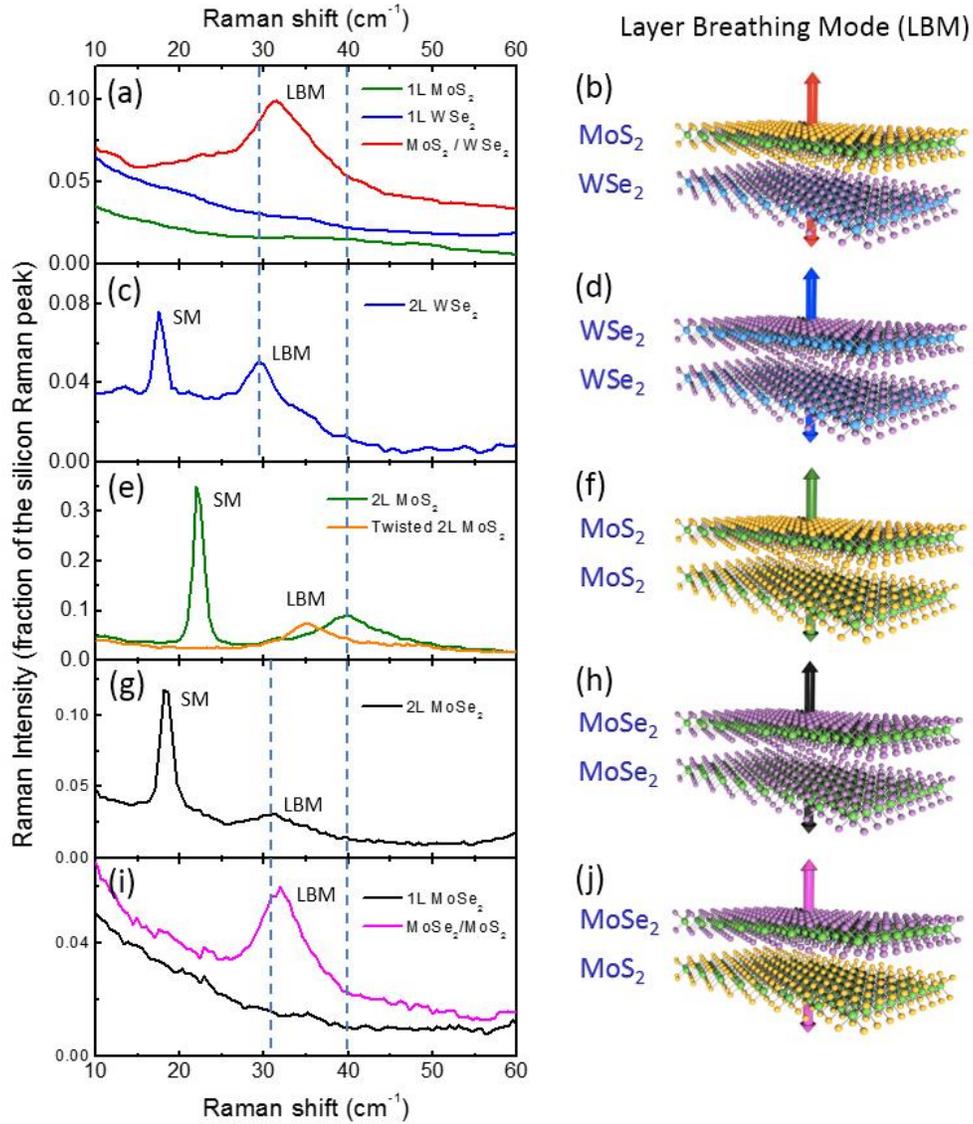

Figure 2. Low-frequency Raman spectra (left column) and schematics of layer breathing mode (LBM) vibrations (right column) for (a-b) the $MoS_2/WSe_2$ heterobilayer, 1L $MoS_2$ and 1L $WSe_2$; (c-d) Bernal-stacked 2L $WSe_2$; (e-f) Bernal-stacked and twisted 2L $MoS_2$; (g-h) Bernal-stacked 2L $MoSe_2$; (i-j) $MoSe_2/MoS_2$ heterobilayer and 1L $MoSe_2$. All spectra were measured with 532-nm laser excitation, and normalized with the Raman peak of the underlying silicon substrate (at 520 $cm^{-1}$). The shear mode (SM) and layer breathing mode (LBM) are denoted. The dashed vertical lines highlight the LBM position in Bernal-stacked 2L $WSe_2$, 2L $MoS_2$ and 2L $MoSe_2$.

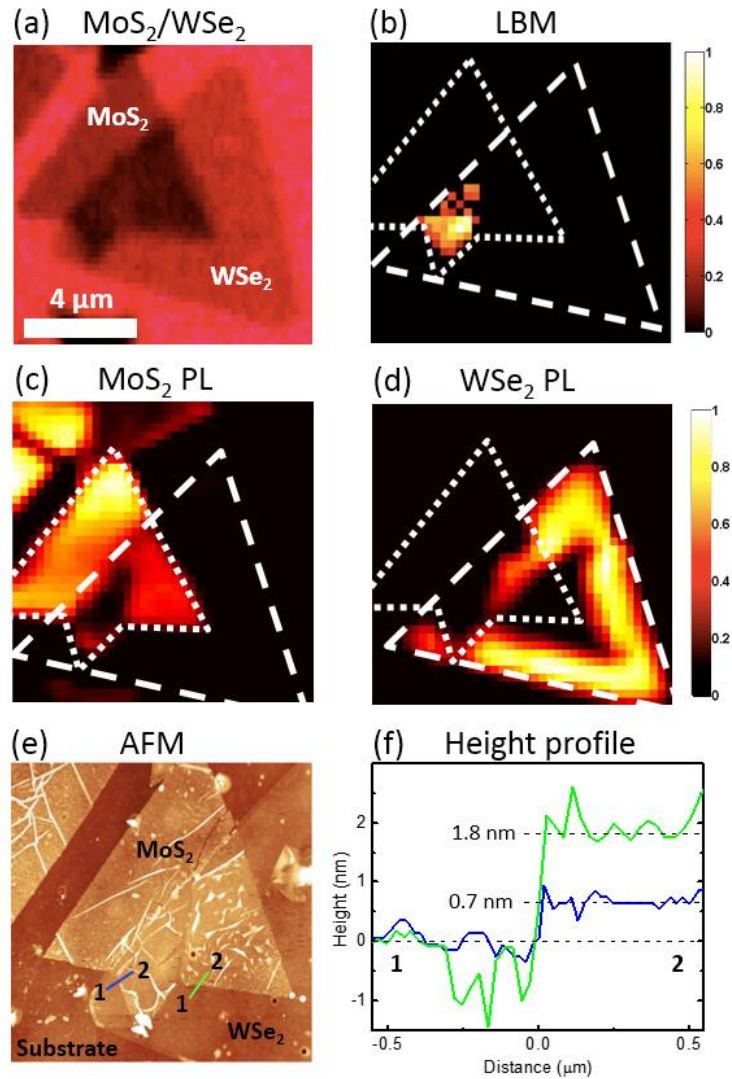

Figure 3. (a) Optical image of an unannealed MoS$_2$/WSe$_2$ heterobilayer. (b) Raman intensity map of the LBM at ~32 cm$^{-1}$ (red curve in Figure 2a). (c) Intensity map of the PL peak at 1.85 eV that corresponds to the main PL band of MoS$_2$ (green curve in Figure 1d). (d) Intensity map of the PL peak at 1.59 eV that corresponds to the main PL band of WSe$_2$ (blue curve in Figure 1d). The PL is enhanced at the edge of the WSe$_2$ monolayer, as also reported in the literature [53]. The short and long dashed lines highlight the regions of 1L MoS$_2$ and 1L WSe$_2$, respectively. (e) AFM topographic image. (f) Height profiles of the two cutting lines across the boundary between the 1L WSe$_2$ region and the MoS$_2$/WSe$_2$ region in Panel (e). We use green and blue color to denote the corresponding lines, and the numbers 1 and 2 to denote the corresponding positions.

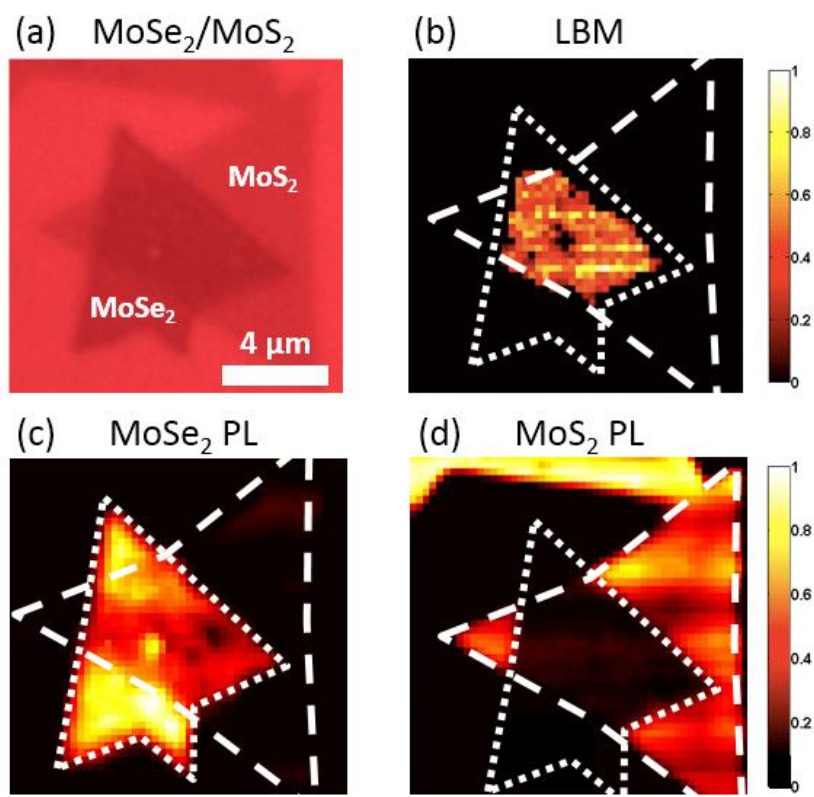

Figure 4. (a) Optical image of an annealed MoSe$_2$/MoS$_2$ heterobilayer. (b) Raman intensity map of the LBM at ~32 cm$^{-1}$ (magenta curve in Figure 2i). (c) Intensity map of the PL peak at 1.56 eV that corresponds to the main PL band of MoSe$_2$ (black curve in Figure 1e). (d) Intensity map of the PL peak at 1.85 eV that corresponds to the main PL band of MoS$_2$ (green curve in Figure 1e). The short and long dashed lines highlight the regions of 1L MoSe$_2$ and 1L MoS$_2$, respectively.

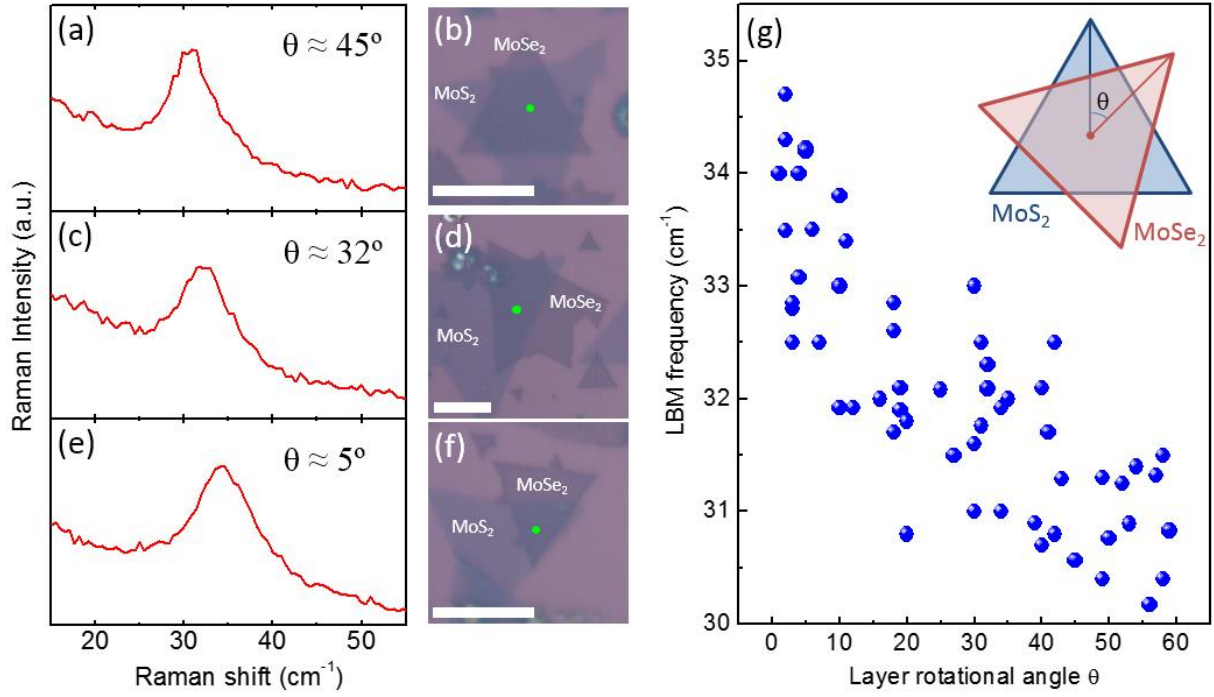

Figure 5. (a-f) Layer-breathing mode (LBM) Raman spectra and the associated optical images of three representative MoSe$_2$/MoS$_2$ heterobilayer samples at different layer rotational angles ($\theta$). The scale bars are 10 μm. The green dots in the optical images denote the position of the laser spot in the measurements. (g) The LBM frequencies of 58 MoSe$_2$/MoS$_2$ heterobilayer samples as a function of the layer rotational angle ($\theta$). The inset shows the schematic rotational angle between single-crystal MoS$_2$ and MoSe$_2$ monolayers.

|  | MoS$_2$/WSe$_2$ | MoSe$_2$/MoS$_2$ | Twisted 2L MoS$_2$ | 2L MoS$_2$ | 2L MoSe$_2$ | 2L WSe$_2$ |
|---|---|---|---|---|---|---|
| LBM frequency (fwhm) (cm$^{-1}$) | 30-35 (8) | 30-35 (5.5) | 32-39 (6.5) | 39.5 (9.5) | 30.5 (5) | 29.5 (3.5) |
| SM frequency (fwhm) (cm$^{-1}$) | --- | --- | --- | 22 (1.5) | 18.5 (1.5) | 17.5 (1.5) |

Table 1. The frequency and full width at half maximum (fwhm) of the layer-breathing mode (LBM) and shear mode (SM) for different TMD bilayers. The uncertainty for all numbers is ±0.5 cm$^{-1}$.

13